\documentclass[11pt,twoside]{article}

\usepackage{asp2006}

\usepackage{epsf}

\usepackage{psfig}

\usepackage{lscape}

\begin{document}

\title{The PSU/TCfA Search for Planets Around Evolved Stars:\\  \boldmath $V \sin i$ Measurements for Slow Rotating F-K Giants.}

\begin{quote}

Grzegorz Nowak$^1$ and Andrzej Niedzielski$^{1,2}$

{\itshape $^1$Torun Centre for Astronomy, Nicolaus Copernicus University,

ul. Gagarina 11, 87-100 Torun, Poland}

{\itshape $^2$Department of Astronomy and Astrophysics, Pennsylvania State

University, 525 Davey Laboratory, University Park, PA 16802}

\end{quote}

\begin{abstract} 
We present results of our projected rotational velocities ($V \sin i$) measurements of F, G and K giants obtained from the cross-correlation function (CCF) constructed from high signal to noise spectra. We also present the calibration of the HET/HRS cross-correlation function to determine accurate projected rotational velocities $V \sin i$ for slowly-rotating F-K giants.
\end{abstract}

\section*{Introduction and observations}

By fitting a Gaussian to the CCF profile we obtain three free parameters: position of the minimum (which is directly related to the radial velocity), dispersion $\sigma_{obs}$ (related to the surface velocity field) and the cross correlation area (related to the stellar metallicity). If we measure $\sigma_{obs}$ for stars for which accurate $V \sin i$ were determined by Fourier analysis or from rotational periods, we may build a calibration of the CCF technique to determine accurate $V \sin i$.

The observational material and reduction are described in Niedzielski \& Wolszczan (this volume).

\section*{The \boldmath $\sigma_{0}$ - $V \sin i$ calibration for HET/HRS and preliminary resulsts}

We obtained  the CCF by cross correlating the high S/N blue spectra with a numerical mask defined as a sum of delta-functions centered on the rest wavelengths of selected lines (since the mask is a mathematical function it does not add noise to the data). After computing the CCFs  for every order, they are adjusted to a common reference frame and co-added to get the final normalized CCF for which the $FWHM$ is measured. The dispersion ($\sigma_{obs}$) is related to $FWHM$ as: $FWHM = 2 \sqrt{2 \ln(2)} \sigma_{obs}$.

\citet{BandM1984} have shown that the width of the CCF ($\sigma_{obs}$) is related to the $V \sin i$ by a function of a form of: $V \sin i = A \sqrt{\sigma_{obs}^{2} - \sigma_{0}^{2}}$, where $A$ is a constant coupling the differential broadening of the CCFs to the $V \sin i$ of a star, and $\sigma_{0}$ is the width of the CCF of an non-rotating star of the same spectral type and luminosity and is related to the measured width of the CCF ($\sigma_{obs}$) by the following formula: $\sigma_{obs}^{2} = \sigma_{0}^{2} + \sigma_{rot}^2$. The measured width of the CCF of a star ($\sigma_{obs}$) results from several broadening mechanisms: gravity, effective temperature, magnetic field, metallicity, and of course the rotation. $\sigma_{rot}^{2}$ is the rotational broadening and $\sigma_{0}^{2}$ is responsible for all other broadening mechanisms (including the instrumental profile) except rotation. $\sigma_{0}$ is a critical parameter. It is a function of the color of a star and we may obtain its dependence on $(B-V)$ by using slowly-rotating calibrator stars. To determine the $\sigma_{0}$ vs. $(B-V)$ relation we used 16 slow rotators with accurate projected rotational velocities measured in various papers, preferably from by \citet{Gray1989}, \citet{Fekel1997} and \citet{deMandM1999}. For every of these stars we determined $\sigma_{0}$ using formula from \citet{BandM1984} and assuming  the constant $A = 1.9$ following the \citet{Queloz1998} and \citet{Melo2001}. We carried out a least-squared fit to the  data by the analytical function $\sigma_{0} = a_{2} (B-V)^{2} + a_{1} (B-V) + a_{0}$ which  yields the following calibration: $\sigma_{0} = 15.592(B-V)^{2} - 26.753(B-V) + 14.559$.

\begin{table}[!h]

\caption{HET/HRS $V \sin i$ for several stars from our survey and their comparison with previous work: (a) \citet{FandV2005}, (b) \citet{DaSilva2006}.}

\label{HETHRSresults}

\smallskip

\begin{center}

{\small

\begin{tabular}{llcccc}

\tableline

\noalign{\smallskip}

Name & Spectral type & $B-V$ & $\sigma_{obs}$ & $V\:sin\,i$ & $V\:sin\,i$\\

 & & & & HET/HRS & other\\

\noalign{\smallskip}

\tableline

\noalign{\smallskip}

HD 17092 & K0 & 1.000 & 3.437 & 0.98 & -\\

HD 38529 & G4V  & 0.773 & 3.551 & 2.94 & $3.90^{a}$\\

HD 118203 & K0 & 0.810 & 3.896 & 4.44 & $4.70^{b}$\\

HD 10697 & G5IV & 0.860 & 3.288 & 2.17 & $2.48^{a}$\\

HD 88133 & G5IV & 0.860 & 3.167 & 1.38 & $2.17^{a}$\\

HD 75732 & G8V & 0.869 & 3.136 & 1.07 & $2.46^{a}$\\

HD 95296 & K0 & 1.000 & 3.484 & 1.46 & -\\

HD 77819 & G5 & 0.860 & 3.845 & 4.37 & -\\

BD+57 114 & G4V & 0.940 & 3.720 & 3.64 & -\\

\noalign{\smallskip}

\tableline

\end{tabular}

}

\end{center}

\end{table}

In table \ref{HETHRSresults} we present projected rotational velocities for several stars from our survey (the typical error of our measurements is about 1.5 km/s). It is clear, that our measurements are in good agreement with previous determinations.

\acknowledgements We thank the HET resident astronomers and telescope operators for cooperation. We acknowledge the financial support from the MNiSW through grant 1P03D 007 30. GN is a recipient of a graduate stipend of the Chairman of the Polish Academy of Sciences.



\begin{thebibliography}{}


\bibitem[Benz \& Mayor(1984)]{BandM1984}

Benz, W., \& Mayor, M. 1984, A\&A, 138, 183


\bibitem[Da Silva et al.(2006)]{DaSilva2006}

Da Silva, R, et al. 2006, A\&A, 446, 717


\bibitem[de Medeiros \& Mayor(1999)]{deMandM1999}

de Medeiros, J.R., \& Mayor, M. 1999, A\&AS, 139, 433


\bibitem[Fekel(1997)]{Fekel1997}

Fekel, F.C. 1997, PASP, 109, 514


\bibitem[Fischer \& Valenti(2005)]{FandV2005}

Fischer, D. A., \& Valenti, J. 2005, ApJ, 622, 1102


\bibitem[Gray(1989)]{Gray1989}

Gray, D.F. 1989, ApJ, 347, 1021


\bibitem[Melo et al.(2001)]{Melo2001}

Melo, C. F. H., Pasquini, L., \& de Medeiros, J. R. 2001, A\&A, 375, 851


\bibitem[Queloz et al.(1998)]{Queloz1998}

Queloz, et al. 1998, A\&A, 335, 183

\end{thebibliography}
\end{document}